\documentclass[a4paper]{jpconf}
\usepackage{graphicx}
\usepackage{amsmath}
\usepackage{hyperref}

\newcommand{\ttbar}{\ensuremath{\mathrm{t\bar{t}}}}
\newcommand{\mumu}{\ensuremath{\mathrm{\mu\mu}}}
\newcommand{\ee}{\ensuremath{\mathrm{ee}}}
\newcommand{\emu}{\ensuremath{\mathrm{e}\mu}}
\newcommand{\mue}{\ensuremath{\mu \mathrm{e}}}

\newcommand{\stat}{\ensuremath{\mathrm{stat.}}}
\newcommand{\syst}{\ensuremath{\mathrm{syst.}}}
\newcommand{\lumin}{\ensuremath{\mathrm{Lumi.}}}
\newcommand{\lumi}{{\mathcal{L}}}
\newcommand{\TeV}{\ensuremath{\mathrm{\,TeV}}}
\newcommand{\GeV}{\ensuremath{\mathrm{\,GeV}}}

\newcommand{\pb}{\ensuremath{\mathrm{\,pb}}}
\newcommand{\fb}{\ensuremath{\mathrm{\,fb}}}

\newcommand{\ifb}{\ensuremath{\mathrm{\,fb^{-1}}}}
\newcommand{\xsec}{\ensuremath{\sigma_{\ttbar}}}
\newcommand{\MET}{\ensuremath{\not \!\! E_T}}
\renewcommand{\pt}{\ensuremath{\mathrm{p_T}}}
\newcommand{\mt}{\ensuremath{\mathrm{m_t}}}

\begin{document}
\title{Measurement of the inclusive $\ttbar$ cross section in the LHC}

\author{Javier Brochero$^{1}$ on behalf of the CMS and ATLAS Collaborations}

\address{$^{1}$Instituto de F\'isica de Cantabria (IFCA-UC). Av. Los Castros, S/N Ed. Juan Jorda, Santander, Spain}

\ead{\href{mailto:javier.brochero.cifuentes@cern.ch}{javier.brochero.cifuentes@cern.ch}}

\begin{abstract}
This document presents recent results of inclusive top-quark pair production cross section measurements at $7$ and $8\TeV$. 
The results are obtained analyzing the data collected by the CMS and ATLAS detectors at the LHC accelerator.   
Studies are performed in the dilepton channel, where the smallest uncertainty is reached, with different approaches.
The most precise results of both experiments are combined and confronted with the most precise theoretical calculation (NNLO-NNLL).
\end{abstract}

\section{Introduction}

The top quark, the heaviest fundamental particle with a mass about $173.3\GeV$ \cite{mt_world}, plays an important role in the study of the electroweak symmetry breaking (Higgs boson) as well as in the search of physics beyond the standard model {(BSM)}. 
Moreover, the production of top quark anti-quark pairs is one of the main backgrounds in many of the processes related with the standard model (SM) and BSM, and it is crucial to measure its production cross section with very high precision. 

The Large Hadron Collider (LHC) has been in operation since 2009, producing proton-proton collisions with a center of mass energy of $7\TeV$ until 2011 and $8\TeV$ in 2012.
This document presents the most recent measurements of the inclusive $\ttbar$ cross section with data collected using the ATLAS \cite{ATLASDetector} and the CMS \cite{CMSDetector} detectors.
 
The most recent theoretical predictions for the top quark pair production cross section ($\xsec$) are $\xsec^\text{NNLO+NNLL}(8\TeV)= 245.9\pm8.4\,\text{(scale)}\, \pm 11.3\,\text{(PDF)}\pb$ and $\xsec^\text{NNLO+NNLL}(7\TeV)= 172.0\pm5.8\,\text{(scale)}\, \pm 8.8\,\text{(PDF)}\pb$ \cite{Czakon2013} for a top-quark mass of $\mt=173.3\GeV$. 
According to the SM, top quarks decay into a W boson and a b quark almost 100\% of the times. 
This leads to final states with two W bosons and two jets coming from the b quark fragmentation.
When both W bosons decay leptonically, the event contains two high momentum leptons with opposite charge, two undetected neutrinos which are measured as missing energy in the plane transverse to the beam axis ($\MET$), and at least two jets, where two of them originate from b quarks.  
The dilepton channel is optimal to measure the inclusive $\ttbar$ cross section due to the small background contribution from other SM processes. As will be presented in the following sections, the uncertainty on $\xsec$ measurement in the dilepton channel is dominated by systematic errors.

\section{Simultaneous measurement of $\ttbar$, $W^{+}W^{-}$ and $Z/\gamma^{*}\to\tau\tau$ cross sections at $7\TeV$}
\label{simulATLAS}
\begin{figure}[ht]
\begin{center}
 \includegraphics[width=0.4\textwidth]{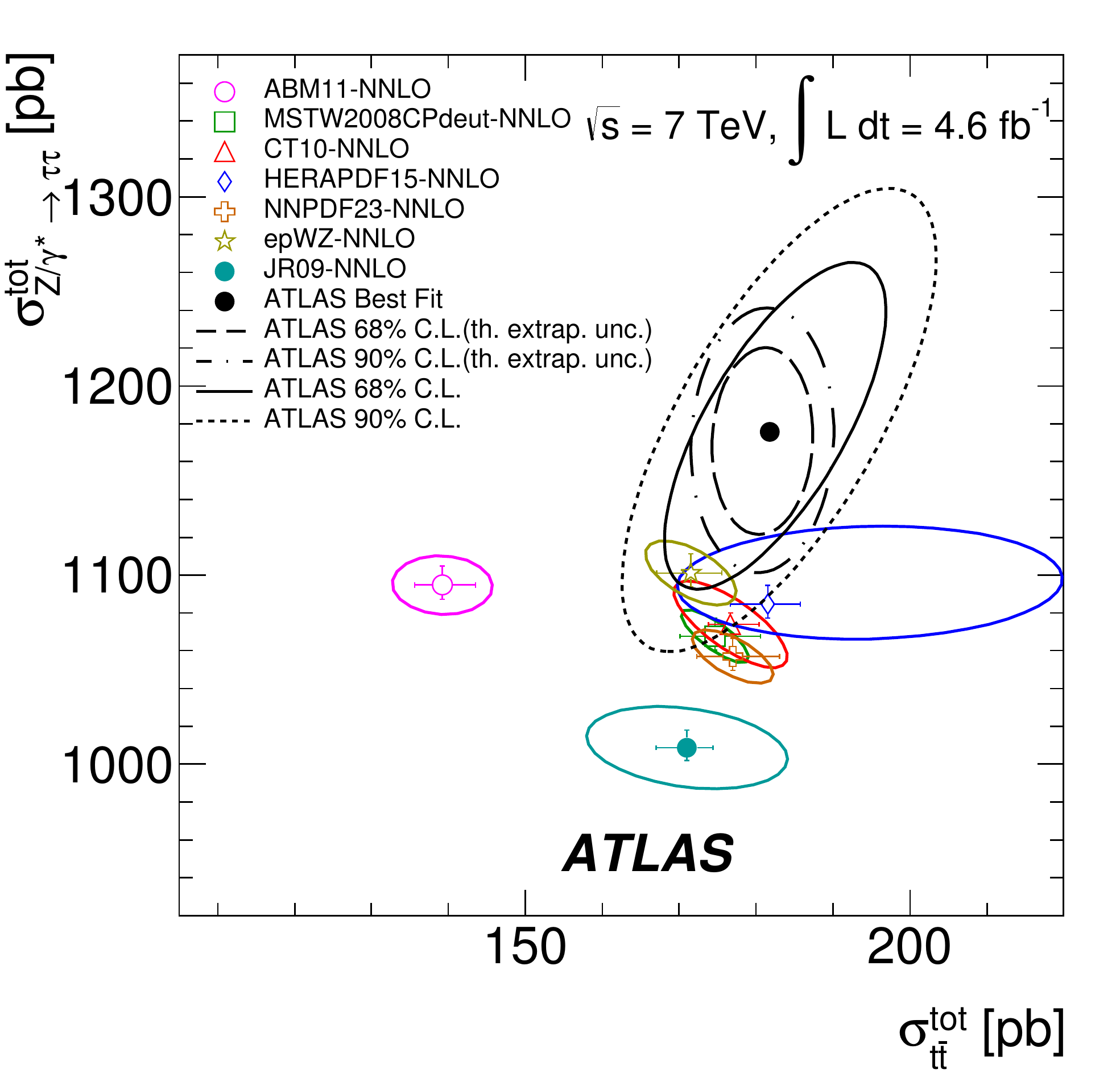} \hspace{1cm}
 \includegraphics[width=0.4\textwidth]{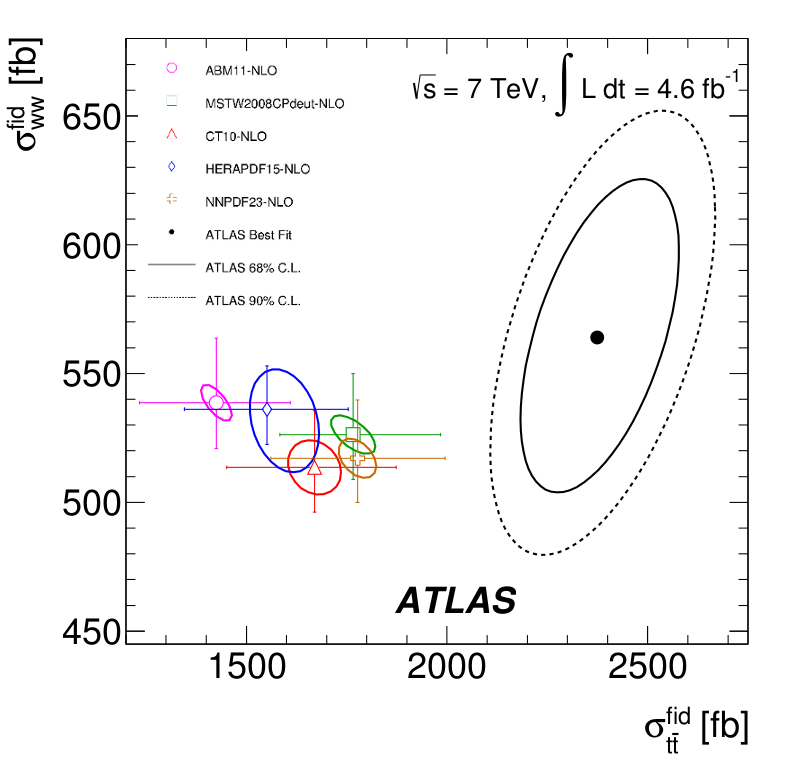}\\
~~~~~~~~~~~~~~~~~~~~~~~~~~~~~~~~~(a)\hfill(b)~~~~~~~~~~~~~~~~~~~~~~~~~~~~~~
\caption{
Contours of the likelihood function as a function of the (a) total and (b) fiducial production cross sections of the $\ttbar$ and Drell-Yan/WW processes \cite{ATLASsimul}. In the case of the total cross section, the result is compared to NNLO available prediction, while the fiducial cross section is compared to the NLO calculation. 
}
\label{ATLAS_Simul}
\end{center}
\end{figure}

This analysis performed by the ATLAS Collaboration \cite{ATLASsimul} exploits the fact that the $\ttbar$, $WW$ and $Z/\gamma^{*}\to\tau\tau$ processes dominate the muon-electron final state.
The simultaneous measurement of all three cross sections is performed at $7\TeV$ using the full 2011 dataset. 
The cross section of these processes is extracted through a fit in a two-dimensional parameter space defined by the missing transverse energy ($\MET$) and the jet multiplicity ($N_{jets}$).
Events from $\ttbar$ production have large $\MET$ and large $N_{jets}$, whereas $WW$ events have large $\MET$ but small $N_{jets}$, and finally, Drell-Yan ($Z/\gamma^{*}\to\tau\tau$) events are characterized by small $\MET$ and small $N_{jets}$.
The results obtained for the $\ttbar$ process are:
\begin{equation}
\begin{aligned}
{\xsec^{\text{total}}}  &= 181.2 \pm 2.8 (\stat) \pm ^{9.7}_{9.5} (\syst) \pm 3.3 (\lumi) \pm 3.3  (\mathrm{{beam}})\pb,\\
{\xsec^{\text{fiducial}}}&=  2730 \pm 40 (\stat) \pm 140 (\syst) \pm 50 (\lumi) \pm 50 (\mathrm{{beam}})\fb,
\end{aligned}
\end{equation}
The fiducial phase space is defined by the lepton cuts applied in the analysis: $\pt^{e(\mu)}>25(20)\GeV$ and $|\eta^{e(\mu)}|<2.5(2.47)$, and it includes the $\tau \to \ell\nu\nu$ decay mode. 
Figure \ref{ATLAS_Simul} shows the best fit for the cross section values with likelihood contours obtained from the simultaneous fit, overlayed with theoretical cross section predictions in the case of the total (NNLO) and fiducial (NLO) phase space.
The theoretical predictions are calculated for different PDF sets. 
A better data-MC agreement is obtained when NNLO calculations are used.

\section{Measurement of the $\ttbar$ production cross section using $\mue$ events with b-tagged jets in pp collisions at $\sqrt{s}= 7$ and $8\TeV$ with the ATLAS detector}
\label{btagATLAS}
\begin{figure}
\begin{center}
  \includegraphics[width=0.4\textwidth]{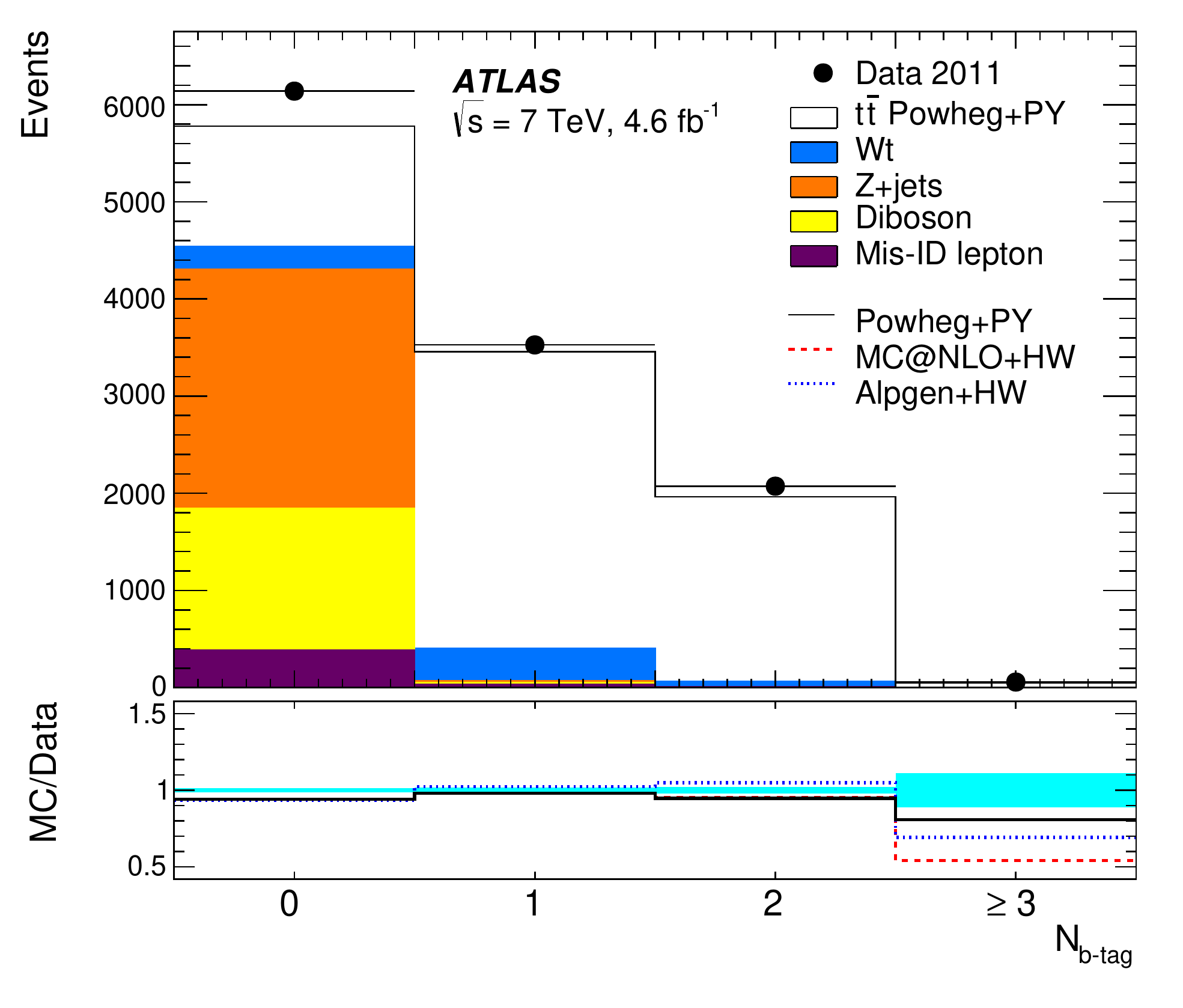}\hspace{1cm}
  \includegraphics[width=0.4\textwidth]{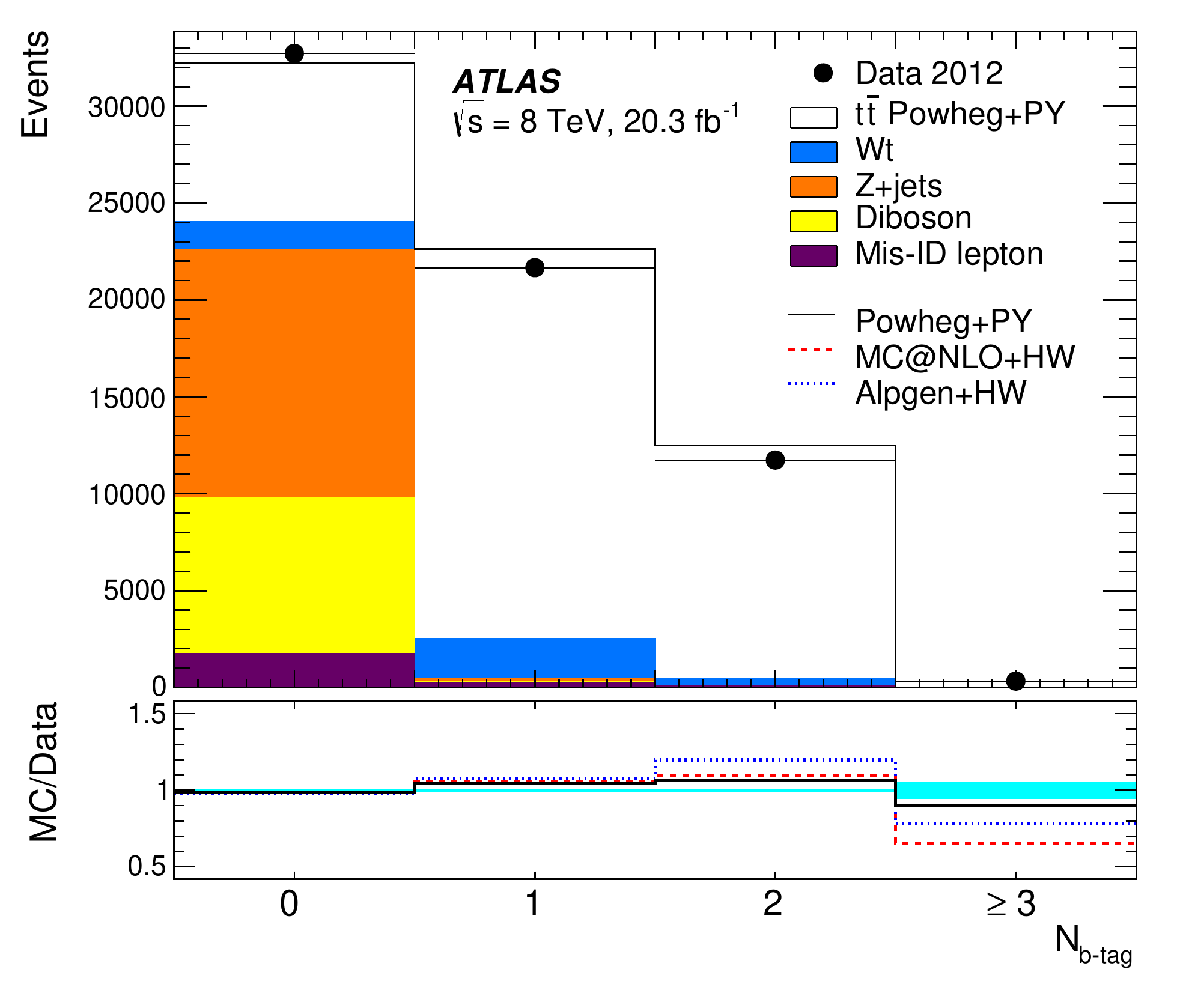}\\
~~~~~~~~~~~~~~~~~~~~~~~~~~~~~~~~~(a)\hfill(b)~~~~~~~~~~~~~~~~~~~~~~~~~~~~~~
  \caption{Distributions of the number of b-tagged jets in $\mue$ events in (a) $\sqrt{s}=7\TeV$ and (b) $\sqrt{s}=8\TeV$ data \cite{ATLASbtag}. 
The data is compared to the expectation from signal and background simulations. }
  \label{ATLAS_btag}
\end{center}
\end{figure}

The following analysis is a dedicated measurement of the $\ttbar$ cross section in the muon-electron decay mode using the full 2011 and 2012 datasets \cite{ATLASbtag}.
The $\ttbar$ production cross section ($\xsec$) is determined by counting the numbers of $\mue$ events with exactly one and exactly two b-tagged jets.
In order to reduce the systematic uncertainties, the $\xsec$ is extracted simultaneously with the combined probability to reconstruct and tag a b-jet from the top-quark decay. 
The obtained cross section results in the $\mue$ channel are shown in Equation \ref{7-8TeVATLAS} with the corresponding ratio ($R_\ttbar^{~}=\xsec{\scriptstyle(8\TeV)}/\xsec{\scriptstyle(7\TeV)}$).  
Figure \ref{ATLAS_btag} shows the b-tagged jet distribution for the $7$ and $8\TeV$ datasets. 

\begin{equation}
\begin{aligned}
\xsec {\scriptstyle(\sqrt{s}=7\TeV)} &= 182.9 \pm 3.1(\stat) \pm 4.2(\syst) \pm 3.6(\lumi) \pm 3.3(\mathrm{beam}) \pb \\
\xsec {\scriptstyle(\sqrt{s}=8\TeV)} &= 242.4 \pm 1.7(\stat) \pm 5.5(\syst) \pm 7.5(\lumi) \pm 4.2(\mathrm{beam}) \pb \\
R_\ttbar^{~} &= 1.326 \pm 0.024(\stat) \pm 0.015(\syst) \pm 0.049(\lumi) \pm 0.001(\mathrm{beam})
\end{aligned}
\label{7-8TeVATLAS}
\end{equation}

\section{$\ttbar$ cross section in the dilepton channel at $\sqrt{s}=8\TeV$ with the CMS detector}
The amount of data studied in this analysis corresponds to $5.3\ifb$ of proton-proton collisions taken by the CMS detector at $8\TeV$ \cite{CMSdilepton}.
The extraction of $\xsec$ is performed in all three dileptonic final states ($\mumu$, $\ee$ and $\mue$, including $W\to \tau+\nu_{\tau}$ as signal). 
The final $\xsec$ result is obtained with a simple and robust cut-based analysis method.
Equation \ref{8TeV_CMS} presents the results for all three channels.
Figure \ref{CMS_dilepton_8TeV} shows the b-tagged jet multiplicity and the $\MET$ distributions for the muon-electron channel, the decay mode with the smallest background contamination.

\begin{equation}
\begin{aligned}
\xsec^{\mumu} &= 244.3 \pm 5.2(\stat) \pm 18.6(\syst) \pm 6.4(\lumi) \pb\\
\xsec^{\ee} &= 235.3 \pm 4.5(\stat) \pm 18.6(\syst) \pm 6.1(\lumi) \pb\\
\xsec^{\mue} &= 239.0 \pm 2.6(\stat) \pm 11.4(\syst) \pm 6.2(\lumi)\pb 
\end{aligned}
\label{8TeV_CMS}
\end{equation}

\begin{figure}
\begin{center}
  \includegraphics[width=0.4\textwidth]{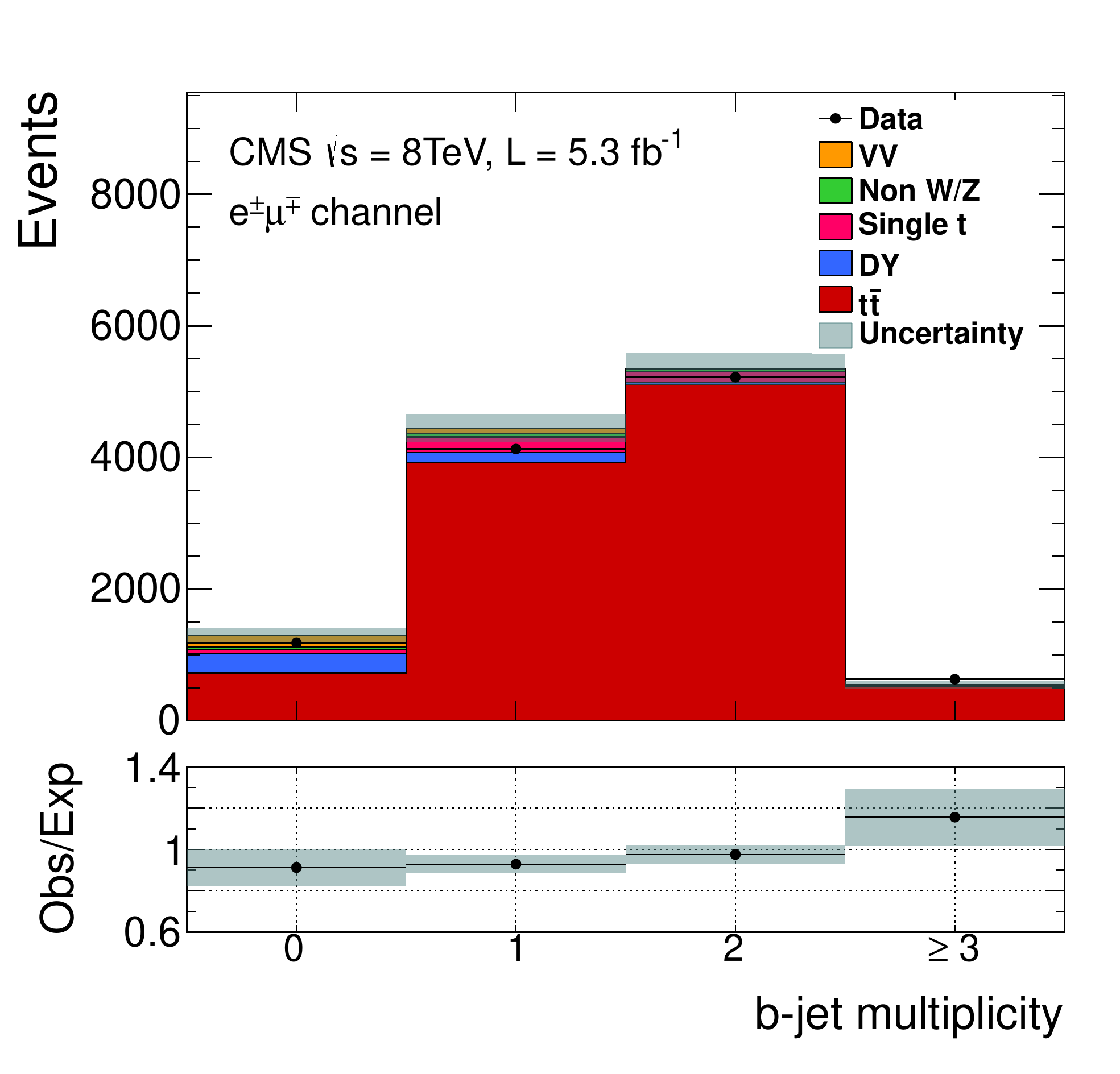}\hspace{1cm}
  \includegraphics[width=0.4\textwidth]{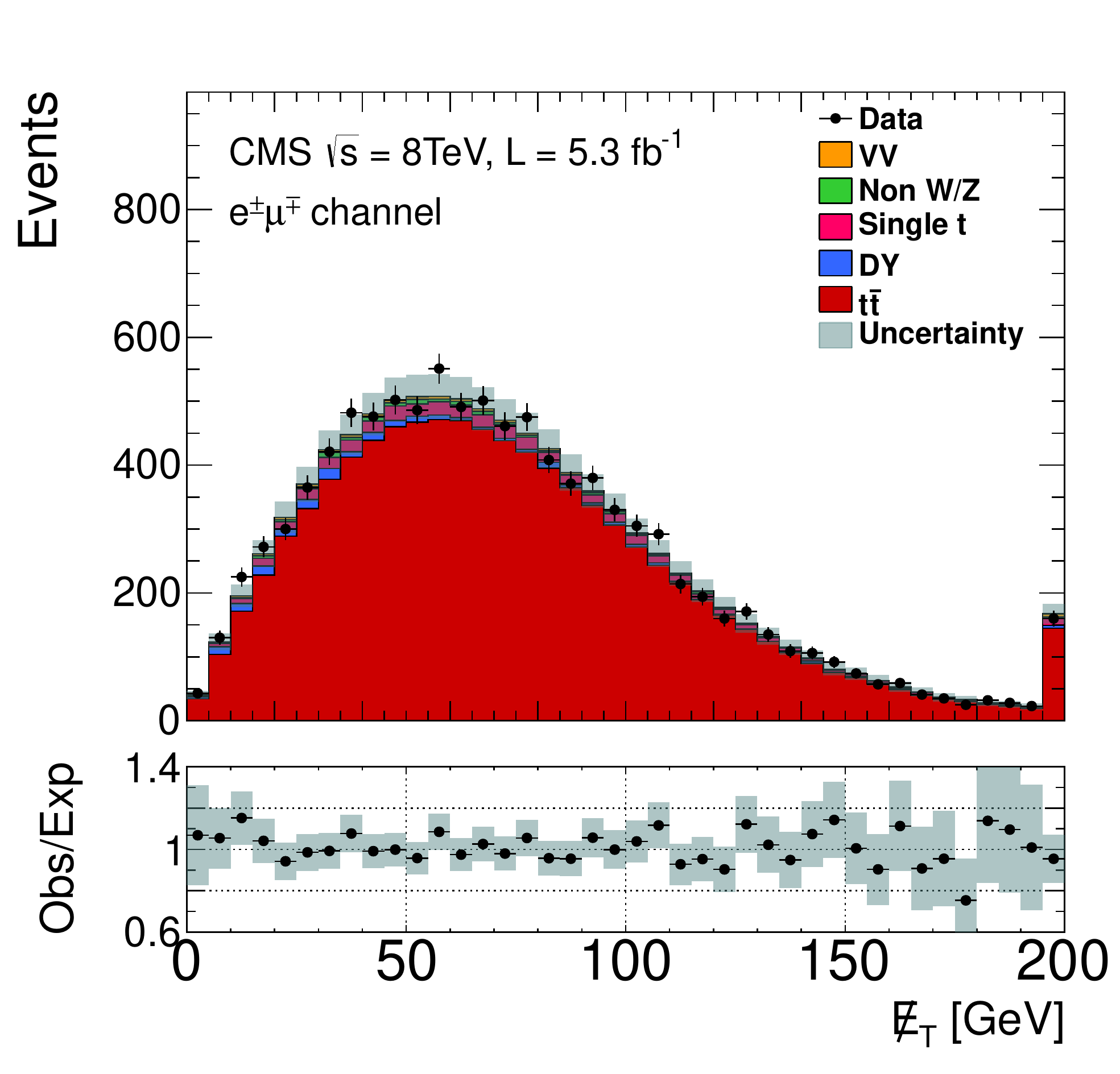}\\
  ~~~~~~~~~~~~~~~~~~~~~~~~~~~~~~~~~(a)\hfill(b)~~~~~~~~~~~~~~~~~~~~~~~~~~~~~~
\caption{(a) b-jet multiplicity and (b) $\MET$ distributions for the muon-electron decay mode \cite{CMSdilepton}. 
The expected distributions for $\ttbar$ signal and background sources are shown by stacked histograms; data are shown by black dots. 
}
  \label{CMS_dilepton_8TeV}
\end{center}
\end{figure}
\section{Measurement of the $\ttbar$ production cross section at $8\TeV$ in dilepton final states containing one $\tau$ lepton}
The CMS collaboration has measured the production cross section of $\ttbar$ pairs by considering dilepton decays where one of the W bosons decays into taus ($\tau\nu_{\tau}$) \cite{CMStaus}.
The identification of the $\tau$ lepton is performed by means of its hadronic decay products.
This decay channel has particular interest because it contributes as a background process in the search for a charged Higgs boson \cite{higgs} ($\ttbar\to H^{+}bW^{-}b^{-}$ with $H^{+}\to\tau\nu_{\tau}$).
The $\ttbar$ cross section is extracted with a cut and count method using the $19.8\ifb$ of $8\TeV$ data collected by the CMS detector in 2012.
Equation \ref{tauCMS} presents the results for the muon-tau ($\mu\tau$) and electron-tau ($e\tau$) channels.
Figure \ref{CMS_dilepton_8TeV} shows the reconstructed $m_t$ and the b-tagged jet multiplicity distributions for the two channels combined.

\begin{equation}
  \begin{aligned}
    \xsec (e\tau)    &= 255 \pm 4(\stat) \pm 24(\syst) \pm 7(\lumin) \pb\\
    \xsec (\mu\tau)  &= 258 \pm 4(\stat) \pm 24(\syst) \pm 7(\lumin) \pb
  \end{aligned}
\label{tauCMS}
\end{equation}

\begin{figure}
\begin{center}
  \includegraphics[width=0.4\textwidth]{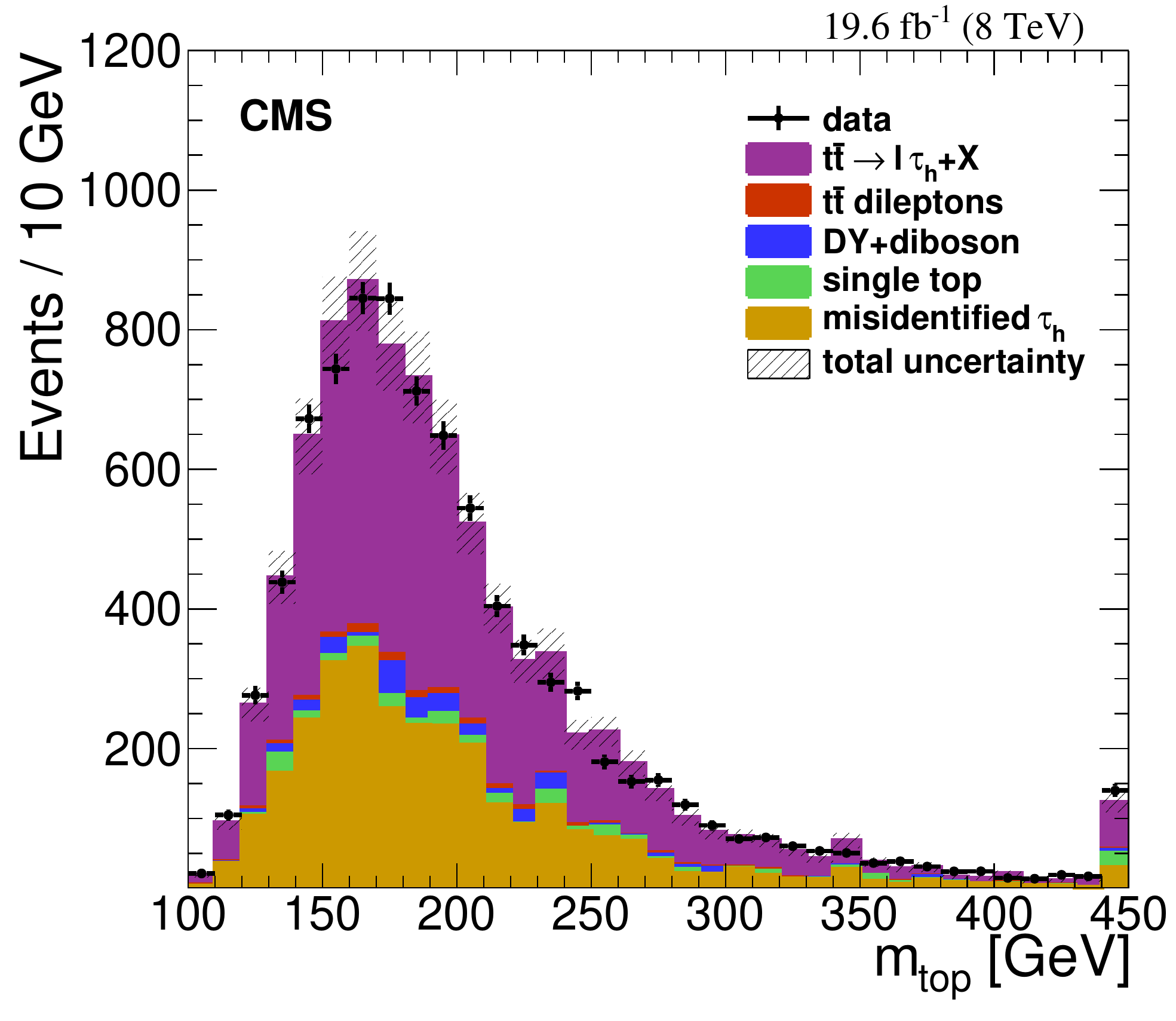}\hspace{1cm}
  \includegraphics[width=0.4\textwidth]{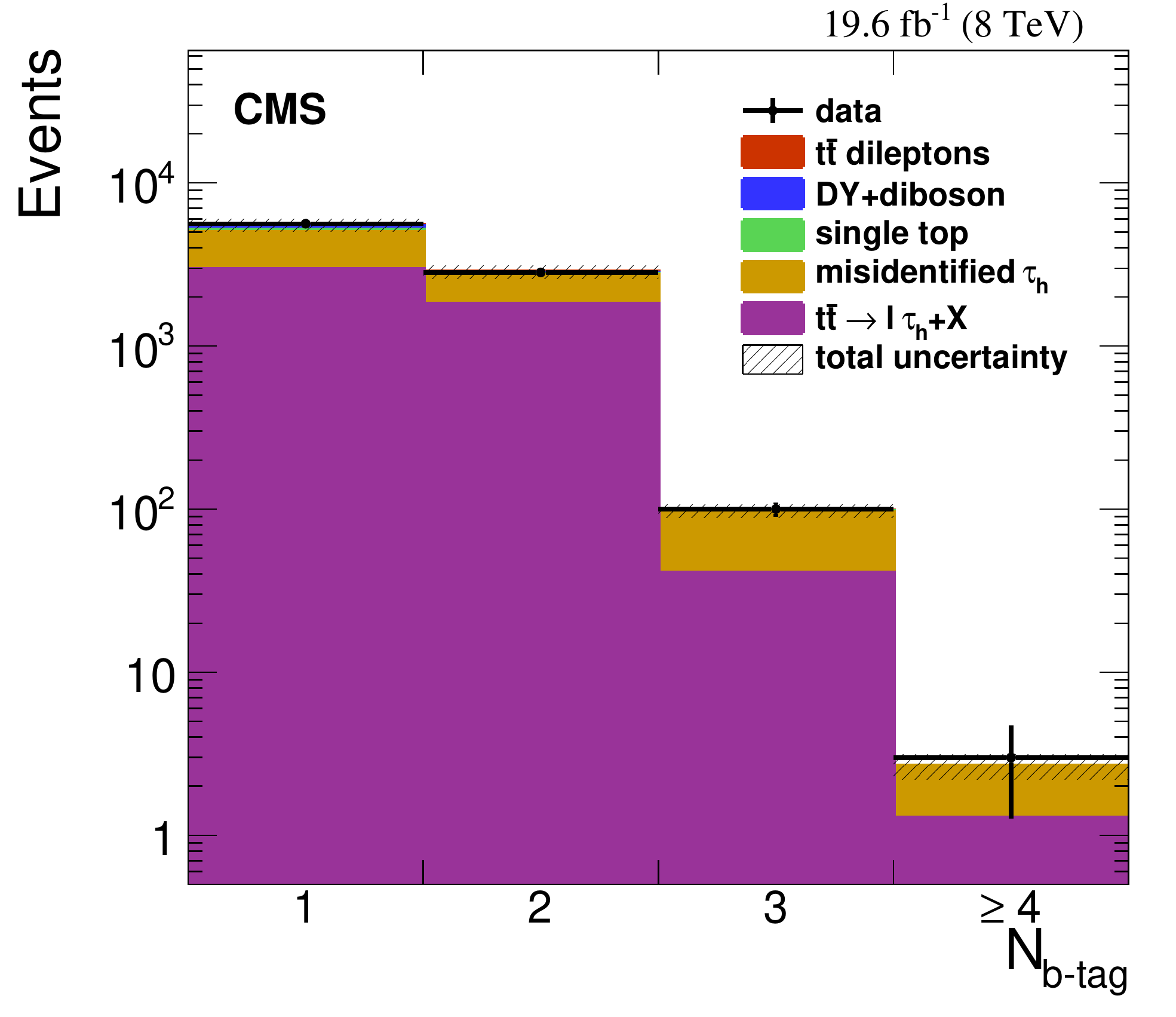}\\
~~~~~~~~~~~~~~~~~~~~~~~~~~~~~~~~~(a)\hfill(b)~~~~~~~~~~~~~~~~~~~~~~~~~~~~~~
  \caption{Distribution of (a) the reconstructed top-quark mass ($m_t$) and (b) the b-tagged jet multiplicity for the $\e\tau$ and $\mu\tau$ channels combined \cite{CMStaus}.
Data (points) are compared with the sum of signal and background yields.
}
  \label{CMS_taus_8TeV}
\end{center}
\end{figure}
\section{Conclusions}
The LHC has a very strong program in the measurement of the inclusive top quark pair production cross section with its two main detectors: ATLAS and CMS. 
The results presented in this document have reached an unprecedented  precision of $\sim 4.5\%$ challenging the current theoretical predictions at NNLO+NNLL.
Figure \ref{CMS_ATLAS_summary} shows the most precise measurements performed by CMS and ATLAS detectors at $7$ and $8\TeV$ compared with the last theoretical calculations.
All results are found to be in agreement with the standard model prediction. 

\begin{figure}
\begin{center}
\includegraphics[width=0.76\textwidth]{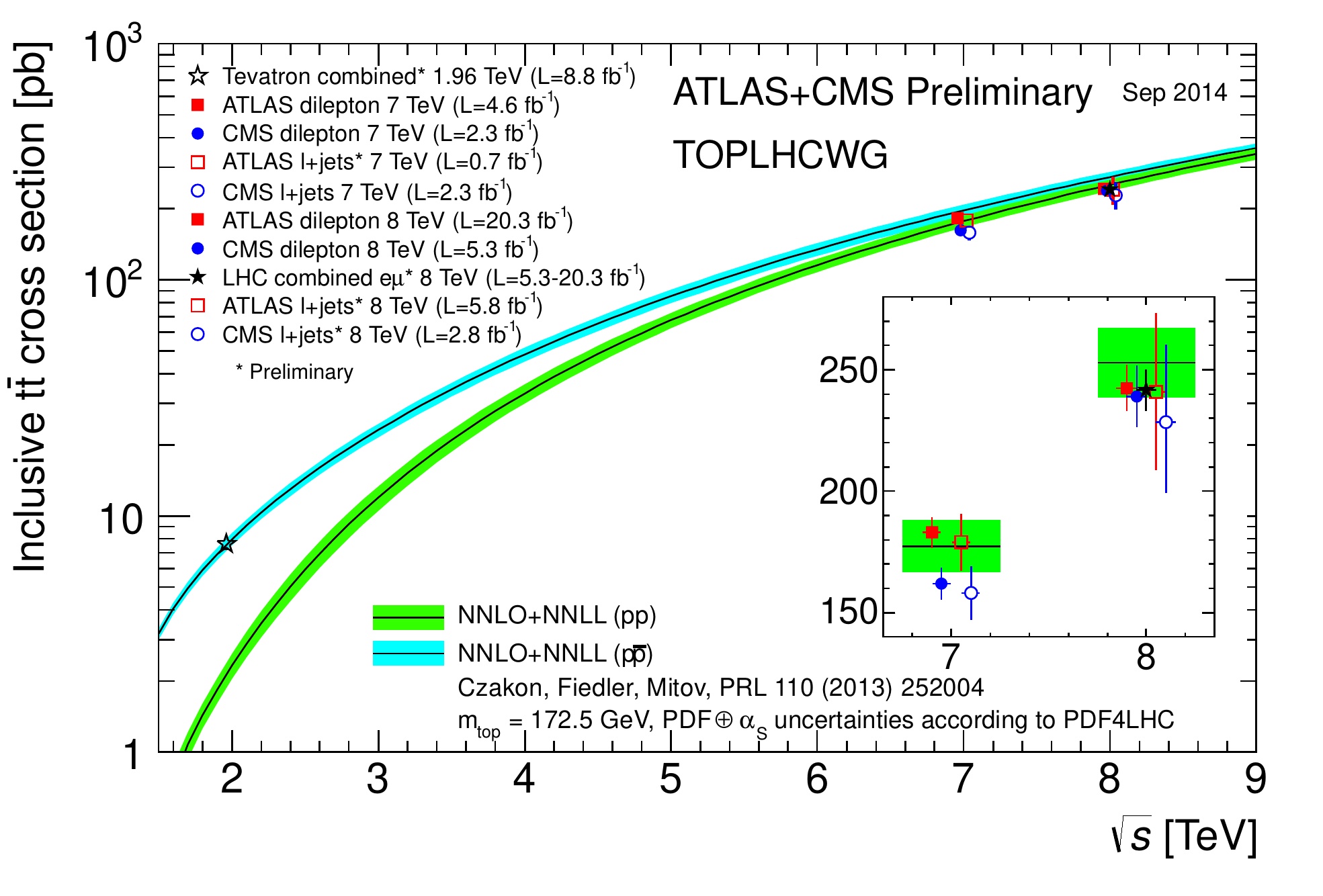}
  \caption{Summary of LHC measurements of the inclusive top-quark pair production cross-section as a function of the center-of-mass energy ($\sqrt{s}$) compared to the NNLO QCD calculation complemented with NNLL resummation. The measurements and the theory calculation is quoted at $m_t=172.5\GeV$. }
  \label{CMS_ATLAS_summary}
\end{center}
\end{figure}

\end{document}